\documentclass[12pt,a4paper]{article}
\usepackage{authblk}
\usepackage{amsthm}
\usepackage[english]{babel}
\usepackage{graphicx}
\usepackage[latin1]{inputenc}
\usepackage{verbatim}
\usepackage{amsfonts}
\usepackage{amsmath}
\usepackage{amssymb}
\usepackage[T1]{fontenc}
\usepackage{color}
\usepackage{epsfig}
\usepackage{natbib}
\usepackage{url}
\usepackage[bookmarksnumbered=true, colorlinks=true, citecolor=blue]{hyperref}
\date{}
\newtheorem{ten}{Tenet}

\title{The Physics and Metaphysics of Pure Shape Dynamics}

\author[a]{Antonio Vassallo
\thanks{\href{mailto:antonio.vassallo1977@gmail.com}{antonio.vassallo1977@gmail.com}}}
\author[a,b]{Pedro Naranjo
\thanks{\href{mailto:pnpfisica@gmail.com}{pnpfisica@gmail.com}}}
\author[c]{Tim Koslowski
\thanks{\href{mailto:t.a.koslowski@gmail.com}{t.a.koslowski@gmail.com}}}
\affil[a]{\normalsize Faculty of Administration and Social Sciences, Warsaw University of Technology, Plac Politechniki 1, 00-661 Warsaw, Poland}
\affil[b]{\normalsize Faculty of Philosophy, University of Warsaw, Krakowskie Przedmie\'scie 3, 00-047 Warsaw, Poland}
\affil[c]{\normalsize University of W\"urzburg, Theoretical Physics II, Campus Hubland Nord, Emil-Hilb-Weg 22, 97074 W\"urzburg, Germany}

\begin{document}

\maketitle

%\begin{center}
\noindent To appear in A. Vassallo (Ed.), \emph{The Foundations of Spacetime Physics: Philosophical Perspectives}. Routledge.\\
The review process for this chapter was handled by Andrea Oldofredi (University of Lisbon).
%\end{center}

\pdfbookmark[1]{Abstract}{abstract}
\begin{abstract}
The goal of this essay is twofold. First, it provides a quick look at the foundations of modern relational mechanics by tracing its development from Julian Barbour and Bruno Bertotti's original ideas until present-day's pure shape dynamics. Secondly, it discusses the most appropriate metaphysics for pure shape dynamics, showing that relationalism is  more of a nuanced thesis rather than an elusive one. The chapter ends with a brief assessment of the prospects of pure shape dynamics in light of quantum physics.\\

\textbf{Keywords}: Pure shape dynamics, relationalism, shape space realism, monism, ontic structural realism.
\end{abstract}

\section{Introduction}\label{sec:1}

 The philosophical debate regarding the nature of space and time is one of the most long-lived disputes in the history of Western thought. The debate is usually presented in terms of the clash between \emph{substantivalism} and \emph{relationalism} (see, e.g., \citealp{483a,483b}). In a nutshell, the doctrine of substantivalism maintains that spatial and temporal structures are ontologically self-subsistent---i.e., their existence is not parasitic upon the existence of matter. Relationalism, on the other hand, denies this claim.\footnote{There is obviously more to this oversimplified story. \citet[][chapter 1, section 3]{118}, for example, distinguishes between an ontological and a semantic version of the anti-substantivalist claim, and points out that the logical relation between the anti-substantivalist claim and the commitment to the only existence of relative motions is not that straightforward. Here we lump together these fine-grained theses in a unique coarse-grained one, since this level of accuracy is sufficient to drive our point home.} Although the substantivalism/relationalism controversy has been pronounced dead several times, especially as far as physics is concerned \citep{58}, it still seems to be alive and kicking. Furthermore, the debate has long transcended beyond the metaphysics of physics, branching out into the philosophy of mathematics and language (cf. the chapters by Cheng \& Read and Bigaj in this volume). The present chapter aims to contribute to the debate by providing a quick yet up-to-date characterization of relationalism and its implementation in a modern physical theory of motion. The utility of such a characterization is evident: For example, \cite{118} laments the fact that relationalism is an elusive doctrine, not least because:
 
\begin{quote}
[T]here is no relationist counterpart to Newton's Scholium, the \emph{locus classicus} of absolutism. Leibniz's correspondence with Clarke is often thought to fill this role, but it falls short of articulating a coherent relational doctrine and it even fails to provide a clear account of key points in Leibniz's own version of relationism [...]\\ %(there is, for example, no mention of Leibniz's reaction to Newton's bucket experiment).\\
(\emph{ibid.}, p. 12)
\end{quote}

The above remark highlights a quite uncontroversial historical fact. Substantivalism is a philosophical doctrine with a clear ``birth date'': Once Newton laid down his Scholium, substantivalism became a well-established stance, with a solid physical backbone. The same thing cannot be said of relationalism. The construction of a philosophically as well as physically sound relational framework has taken centuries, and the work of many philosophers and physicists (see \citealp{77}, for a historical overview of the conceptual development of relationalism). But even taken for granted that relationalism learned to walk when substantivalism was already running, this does not automatically imply that relationalism cannot be as conceptually clear as substantivalism. In fact, modern relationalism is not an elusive doctrine but, rather, a conceptually nuanced one.\footnote{One may object that we are just being evasive on relationalism's elusiveness by preferring from the outset a particular strand of the doctrine (what \citealp{464}, calls \emph{Machian} relationalism) over the others. A defence of our choice would require a paper on its own. For the time being, we just submit that Machian relationalism is the only stance so far codified in a well worked out theoretical framework, whereas the other variants are better seen as ``interpretations'' of standard, non-relational theories such as Newtonian mechanics.} The following sections will substantiate this claim.

We will start with a brief, non-technical overview of modern relational physics, with a special emphasis on the evolution of the approach in relation to the technical implementation of the main relationalist tenets (section \ref{sec:2}). Having established the latest results in the quest for a totally relational theory, we will proceed with considering the possible metaphysical morals compatible with them (section \ref{sec:3}). The discussion will make it apparent that the relational tenets go along with a surprisingly wide array of metaphysical positions, thus highlighting how a relationalist metaphysics is conceptually much richer than the simple ontological denial of space and time and the countenance to relative motion. The chapter will close with some speculations about the future of relational physics, especially in light of quantum physics.

\section{From Best-Matching to Pure Shape Dynamics}\label{sec:2}

The goal of this section is to provide a concise portrait of the main recent developments that have occurred in the relational physics program towards the articulation of a ``coherent relational doctrine''---paraphrasing Earman's quotation above. 

%The goal of this section is to provide a concise historical portrait of the main developments that have occurred in the relational physics programme dubbed \emph{Shape Dynamics} (SD), tracing its roots in the attempt to give a mathematically sound and robust formulation of Mach's principle.

\subsection{The Modern Relationalist Tenets}
\label{Mach}

The birth of modern relational dynamics can be traced back to three seminal papers by Julian Barbour and Bruno Bertotti \citep{722, 84, 83}, which led to the implementation of what is today known as the \emph{Mach-Poincar\'e principle}:

\begin{ten}[\textbf{Mach-Poincar\'e principle -- classic version}]
\label{MachP}
Physical, i.e., relational initial configurations and their (intrinsic) first derivatives alone should uniquely determine the dynamical evolution of a closed system.
\end{ten}

The motivation for tenet \ref{MachP} stems from the relationist desire to eliminate redundant structure from physical theories, which typically arises because of the presence of various symmetries in the physical system under scrutiny. This aversion to redundant structure may be dressed with heavy empiricist overtones by claiming that \emph{all structures whose variation amounts to no empirically observable difference should be banned from physics}. Not surprisingly, Newton's space and time are the redundant structures \emph{par excellence} for the relationist due to their dubious causal efficacy and unobservable nature. Indeed, the Mach-Poincar\'e principle follows from two well-known anti-substantivalist theses:

\begin{ten}[\textbf{Spatial relationalism -- classic version}]
\label{MachP2}
Lengths, be they distances or sizes, must be defined relative to physical systems, not spatial points.
\end{ten}

\begin{ten}[\textbf{Temporal relationalism}] 
Temporal structures, such as chronological ordering, duration, and temporal flow, must be defined only in terms of changes in the relational configurations of physical systems. 
\end{ten}

Moreover, the modern relationalist places a particular emphasis on the fact that every measurement simply amounts to the comparison of two physical systems. Formally, this means that only \emph{ratios} of physical quantities carry objective information.\footnote{Since we are here providing a brief overview of modern relationalism, we refrain from commenting further on this point. See \cite{724,723,725}, for a discussion on the possible consequences of comparativism towards physical magnitudes.} Going back to the case of space and time, this comparativist attitude implies that the relationalist should let go of any notion of size. What remains is simply the \emph{shape} of the system.\footnote{Actually, parameters like masses and charges play an important role in the dynamics, even if they only enter as dimensionless ratios. Strictly speaking, then, what remains after symmetries are removed is the shape of the system weighted by said parameters. For simplicity's sake, we shall leave them out of the discussion, since the essential tenets go through regardless.} To illustrate the idea, let us consider 3 particles in Newtonian space, which may be thought of as placed in the vertices of a triangle. Let $x_a$ be the particle positions and $r_{ab}\equiv|x_a-x_b|$ the inter-particle separations. The empirical fact mentioned implies that we should take one of the $r_{ab}$ as a unit of length and compare the remaining two against it, yielding the two ratios which are objective. The upshot of this procedure is the fact that only the shape of the triangle matters, for only two independent angles are needed. We can hence modify tenet \ref{MachP2} to read:

\begin{ten}[\textbf{Spatial relationalism -- modern version}]\label{MachP3}
The only physically objective spatial information of a physical system is encoded in its shape, intended as its dimensionless and scale-invariant relational configuration.
\end{ten}

Mathematically, the procedure of systematically getting rid of redundant structure associated with some symmetry is commonly known as \emph{quotienting out}. Schematically, if $\mathcal{Q}$ is the relevant configuration space of a given system and $\mathcal{G}$ is the symmetry group, the quotienting out yielding the space carrying truly physical information is $\mathcal{Q}_{ss}:=\mathcal{Q}/\mathcal{G}$, whose dimension is simply $\mathrm{dim} (\mathcal{Q}_{ss})=\mathrm{dim}(\mathcal{Q})-\mathrm{dim}(\mathcal{G})$. In the case of $N$ classical particles, $\mathcal{Q}$ is standard configuration space, $\mathcal{G}=\mathsf{Sim(3)}$, the joint group of Euclidean translations $\mathsf{T}$, rotations $\mathsf{R}$ and dilatations $\mathsf{S}$ (called \emph{similarity group}), and $\mathcal{Q}_{ss}=\mathcal{Q}/\mathsf{Sim(3)}$ is referred to as the \emph{shape space} of the system. It is easy to see why such a reduced configuration space is called like this: In this simple scenario, the objective spatial information of the $N$-particle system is encoded in the shape of the $N$-gon defined by the $N$ particles. Hence, in this case, each point on shape space represents an $N$-gon's shape (i.e., an equivalence class of $N$-gons under $\mathsf{T,\,R\,\text{and}\,S}$ transformations).\footnote{In order to strip each particle of any intrinsic identity---which would constitute a non-relational feature, one may add the permutation group to the symmetries to be quotiented out. \label{perm}}

In \cite{83}, the effective implementation of tenet \ref{MachP} in the context of classical particle dynamics is carried out by means of the so-called \emph{best-matching}, namely, a sort of intrinsic derivative that allows one to clearly establish a measure of \emph{equilocality}, i.e., to be able to know when two points are located in the same position at different times provided relational data alone are given. Importantly, Barbour-Bertotti consider the \emph{relative configuration space}, arrived at after quotienting by translations $\mathsf{T}$ and rotations $\mathsf{R}$. The desire to build a scale-invariant theory in compliance with tenet \ref{MachP3} led to \cite{419}, where also dilatations $\mathsf{S}$ are taken into account: It was at this point that shape space entered the modern relational program.\footnote{This is not to say that shape space methods in physics were not known before \cite{419}. To the contrary, the notion of shape space had already had very useful applications in, e.g., molecular physics (see \citealp{449}, for a textbook on the subject.)}  However, the introduction of scale transformations comes at a price: In order to account for dynamics and match empirical data, a non-shape degree of freedom has to be introduced---the ratio of dilatational momenta.\footnote{The dilatational momentum is defined as $D\equiv\sum _a^N \mathbf{r}_a^{\mathrm{cm}}\,\cdot \mathbf{p}^a_{\mathrm{cm}}$. Given its monotonicity, the ratio $D/D_0$, with $D_0$ some arbitrary choice, has arguably been used as a physical time variable (\citealp{712,706,529}).} The upshot is a relational formulation of classical dynamics as far as the system in its entirety is considered, but which becomes the usual Galilean-invariant dynamics when only subsystems are taken into account (more on this at the end of subsection \ref{psd}). This is a clear vindication of Mach's idea that the fully relational character of the dynamics can be attained only when the whole universe is considered.
 
It is illuminating to reflect a bit on this issue, for it is central to any relational description of physics. It was a well-known fact to pioneers of relational approaches that observable initial configurations and their first derivatives alone do not suffice to uniquely determine the evolution of a system. The culprit is the total angular momentum $\mathbf J$. It had been known since Lagrange's work that the value of $\mathbf J$ of a system cannot be obtained from said initial data.\footnote{Poincar\'e was of course well aware of---and bothered by---this fact, and this had a crucial importance in his formulation of what we have called the Mach-Poincar\'e principle above (the \emph{locus classicus} for this discussion is \citealp[][chapter VII]{732}).} One of the remarkable consequences of best-matching is the vanishing of $\mathbf J$ for a \emph{closed} system---like the whole Universe. This, of course, does not prevent subsystems from having a non-zero angular momentum. Thus, by seriously considering the holistic character of relational physics, \emph{closed} systems alone are elevated to genuinely distinct systems. %And best-matching imposes definite constraints on the values --zero-- of both $J$ and $D$ for the whole Universe, the only truly closed system there is, thereby removing the deficiency that had Poincar\'e so much troubled. 

%What strikes as particularly odd is the fact that great Poincar\'e seemed to have missed the widely known fact already in 1902 that the Solar System is not the whole Universe. The former no wonder does exhibit non-zero angular momentum, whereas the latter does not.

%\footnote{In general, if both rotations and dilatations exist, we can form dynamically effective ratios which are alien to shape space, hence jeopardizing the deterministic character of the principle \ref{MachP}, as from the same initial data in shape space, different motions would be allowed.}

The search for a theory of gravity that fully complies with the relational tenets of \cite{419} eventually led to \emph{Shape Dynamics} (SD).\footnote{It is beyond the scope of this short survey to go into the technical details of SD. The interested reader is advised to look at the nice introduction in \cite{720} as well as the comprehensive, yet pedagogical book by \cite{514}.} SD is a theory of conformal 3-geometries, as opposed to standard Einstein's gravity, which is a theory of Riemannian 4-geometries. In other words, the dynamical variables of the theory are only the parts of a Riemannian $3$-metric that determine angles. Note how, also in this case, a conformal $3$-geometry is what is left of a Riemannian $3$-geometry after translations, rotations, and scale degrees of freedom are quotiented out. The core approach of SD represents a conceptual continuity with Barbour and Bertotti's original relational framework in the following sense: Given two configurations (i.e., two conformal $3$-geometries), find the curve that minimises, by means of best-matching, their intrinsic distinctness. Crucially, though, the same caveat mentioned above in the particle case carries over to dynamical geometry: The physical arena of SD is not conformal superspace, as one would have thought for an allegedly scale-invariant theory: An additional degree of freedom must be introduced, which is associated with the total volume of space.\footnote{Mathematically, let us denote $\mathsf{Riem}$ the set of Riemannian $3$-geometries and $\mathsf{Diff}(3)$ the group of spatial diffeomorphisms. Then, $\mathsf{Superspace}=\mathsf{Riem}/\mathsf{Diff}(3)$. Further, let $\mathsf{S}$ be conformal superspace $\mathsf{S}=\mathsf{Superspace}/\text{conformal transformations}$. Finally, the physical space of SD is $\mathsf{S}_V\equiv\mathsf{Superspace}/\text{VPCT}=\mathsf{S}\times\mathbb{R}^+$, where $\text{VPCT}$ is the group of \emph{volume-preserving} conformal transformations and $\mathbb{R}^+$ represents the spatial volume or its conjugate variable, the so-called \emph{York time}.}

An immediate worry surfaces at this point. Given that SD is a theory of conformal $3$-geometries, isn't it the case that it conflicts with tenet \ref{MachP2}/\ref{MachP3} about spatial relationalism? After all, there is nothing in a conformal $3$-geometry that may ground the existence of such a shape as a relational configuration of \emph{something} material. If such a shape is indeed a web of relations, the \emph{relata} seem to be something akin to geometric points. If that is the case, isn't this a sophisticated form of substantivalism in disguise?\footnote{This objection is put forward, e.g., in \cite{726}.} There are two possible replies to this objection. The first is to go for an ``egalitarian'' interpretation of the gravitational field as yet another physical field---more precisely, a massless spin-$2$ field. Under this reading, a conformal $3$-geometry is just a mathematical description of something as material as, say, the electromagnetic field. And, indeed, the modern relationalist has the resources to incorporate entities with an infinite number of degrees of freedom in her framework (SD being the case in point). In the gravitational case, the material \emph{relata} would then be field magnitudes rather than spatial points.\footnote{We note \emph{en passant} that this is one of the reasons that lead Rynasiewicz to claim that the substantivalism/relationalism debate is trivialized in modern physics: The fact that it is possible to switch from a ``spacetime'' to a ``gravitational field'' talk in general relativistic physics without changing the physical gist of the theory is a sign that the dispute is merely verbal (cf. \citealp[][section V]{58}).} The problem with this response is that the ``egalitarian'' interpretation of the gravitational field in general relativistic physics is problematic and hugely controversial (see, e.g., Pitts' chapter in this book to catch a glimpse of the debate). A second response may be to show that matter can be incorporated in SD, thus constituting the ``backbone'' of a conformal $3$-geometry. The problem with this response is that, at the moment, there is no easy way to couple matter to the purely gravitational sector of SD (see \citealp{547}, for a preliminary attempt at doing so). As things stand, this is still an open issue.

\subsection{The Pure Shape Dynamics Program}
\label{psd}

After this brief tour on the development of relational approaches to dynamics, we finally get to the latest refinement of the theory, dubbed \emph{Pure Shape Dynamics} (PSD; see \citealp{729}, for a general technical introduction to the framework). In a nutshell, the qualifier ``Pure'' means that PSD describes \emph{any} dynamical theory exclusively in terms of the intrinsic geometric properties of the \emph{unparametrized} curve $\gamma _0$ traced out by the physical system in shape space $\mathcal Q_{ss}$, hence ensuring that there are no external reference structures nor clock processes necessary to describe $\gamma _0$ in $\mathcal Q_{ss}$. In other words, the PSD project is highly innovative: It aims at rewriting the whole spectrum of theories of dynamics that characterized and will characterize the development of physics in fully relational, scale-invariant terms. For any ``standard'' physical system, be it a Newtonian $N$-particle system, a general relativistic cosmological model, a de Broglie-Bohm particle system, etc., PSD proposes the same recipe: Quotient out the relevant symmetries from the system and then fully geometrize its dynamical development. To further stress this insistence on the intrinsic geometric properties of the curve associated with a physical system, one speaks of its equation of \emph{state}, in contrast to its equation of motion. We may slightly modify tenet \ref{MachP} to match PSD ``intrinsically geometric'' nature as follows:

\begin{ten}[\textbf{Mach-Poincar\'e principle -- modern version}]
\label{MachPSD}
Physical, i.e., relational initial configurations and their (intrinsic) derivatives alone should uniquely determine the dynamical evolution of a closed system.
\end{ten}

The key innovation brought about by tenet \ref{MachPSD} is to consider in general \emph{higher-order} derivatives of the curve, thereby allowing us to describe the dynamics of a physical system in terms of the curve alone, without the need of any additional non-shape parameters as in standard SD. Clearly, there is a sense in which tenet \ref{MachPSD} is \emph{weaker} than tenet \ref{MachP}, for the former requires more initial data than the latter (higher-order versus first derivatives) to describe a dynamical system. This weakening is anything but a drawback from a modern relationalist standpoint since the upshot is the elimination of any non-shape degree of freedom from the dynamical description of a system---thus delivering a truly relational dynamics cast in terms of the degrees of freedom intrinsic to the system only.

Given an already ``quotiented out'' physical system, we shall express the equation of state of the unparametrized curve $\gamma _0$ in its associated shape space $\mathcal Q_{ss}$ as follows:

\begin{equation}
\begin{array}{rcl}
   dq^a&=&u^a(q^a,\alpha _I^a)\,, \\
   d\alpha _I^a &=&A _I^a(q^a,\alpha_I^a)\,,
%   dk&=&K(q,\phi ,k)
   \end{array}
   \label{curve0}
\end{equation}

and demand that the right-hand side be described in terms of dimensionless and scale-invariant quantities, whose intrinsic change is obtained employing Hamilton's equations of motion. In \eqref{curve0}, $q^a$ are points in shape space, namely they represent the universal configurations of the system, $u^a$ is the unit tangent vector defined by the shape momenta $p_a$: 
\begin{equation*}
    u^a\equiv g^{ab}(q)\frac{p_b}{\sqrt{g^{cd}p_cp_d}}\,,
    \label{unittangent}
\end{equation*}
which allows us to define the direction $\phi ^A$ at $q^a$. It is through the unit tangent vector and the associated direction that the shape momenta enter Hamilton's equations, which are in turn used in the intermediary steps leading to the equation of state \eqref{curve0}. Finally, $\alpha _I^a$ is the set of any further degrees of freedom needed to fully describe the system. It is this set $\alpha _I^a$ that includes higher-order derivatives of the curve and spares us of the need of additional non-shape degrees of freedom. For consistency, the elements in $\alpha _I^a$ must exhaust the set of all possible dimensionless and scale-invariant quantities that can be formed out of the different parameters entering a given theory. Note the unifying nature of \eqref{curve0}: In principle, the whole of relational dynamics, classical, relativistic, and quantum, boils down to the dynamical structure encoded in the above equation of state.  

Two major differences with standard SD should be emphasized. Firstly, as already pointed out above, PSD exclusively relies on the intrinsic geometric properties of the dynamical curve in shape space, as given in \eqref{curve0}, whereas standard SD does not share this prominent role of said curve. In particular, this insistence on intrinsic properties is best exhibited by the unparametrized character of the curve in PSD, which, recall, guarantees that no external reference structures nor clock processes are needed to describe $\gamma _0$, the $\alpha _I^a$ alone being responsible for this job. Secondly, the way the two compare configurations: Whereas standard SD relies on best-matching to obtain the dynamical curve, which involves taking two configurations to find the minimal amount of intrinsic change in their shapes, PSD proposes an initial-value formulation, whereby a \emph{single} set of relational initial data in \eqref{curve0} generates the dynamical curve.

As already stressed above, there is an important reason for considering PSD an improvement of the original SD framework. In short, the dynamical evolution as described by PSD does not fundamentally rely on any notion of parametrization whatsoever and, hence, it is a genuinely intrinsic description of a physical system. Remarkably enough, PSD is capable of reproducing known physics despite its decidedly intrinsic nature (see \citealp{729}, for the $E=0$ $N$-body problem; the case of dynamical geometry will be the subject of another paper, currently in preparation). On the other hand, the original formulation of SD needs some monotonically increasing parameter---be it the ratio of dilatational momenta or York time/spatial volume, as discussed earlier---to be defined on a dynamical curve in order to make sense of the physical evolution. Such a parameter, although representing a much weaker structure than Newtonian time, still represents something ``external'' to the system. 

%(see \citealp{729}, for the $E=0$ $N$-body problem; for the case of dynamical geometry, see \citealp{712}, for the structural similarities between the $E=0$ $N$-body problem and $\Lambda =0$ General Relativity in standard SD. These similarities naturally carry over to PSD, whose full implementation is work in progress/presented elsewhere).

This brings us to one of the key aspects of PSD, namely its account of dynamics. Given that we deal with unparametrized curves,
%which means the above account of dynamics is not available. 
the challenge is to find an intrinsic feature of the system that serves the purpose of a physically meaningful labeling of change. To this extent, we shall exploit the idea put forward in \cite{706} that a shape contains structure encoded in stable records.\footnote{Although Barbour and collaborators worked in the context of standard SD, their main results can be easily carried over to PSD, as the ensuing discussion will clarify.} Thus, the evolution is towards configurations (i.e., shapes) that maximize the complexity of the system. The result provides the desired ground for introducing a direction of change, which boils down to the direction of accumulation of the above-mentioned stable records.\footnote{See the exchange between \cite{710} and \cite{711} for a critical discussion of this point.} Clearly, the first thing we must provide is a natural definition of complexity, along with a suitable measure of it, which, to comply with the central tenet of PSD, should be given in terms of the intrinsic geometry of the unparametrized curve in shape space.  

%\todo[inline]{To make sense of a measure of complexity and its growth, the latter being somewhat unsatisfactory in JFT}

Given that the only fully worked-out physical system exhibiting generic formation of stable records is the $N$-body system, we shall use it as an example to motivate our approach to dynamics. It is worth pointing out that promising results come from the vacuum Bianchi IX cosmological model, where a natural candidate exists for a measure of shape complexity in geometrodynamics (see \citealp[][section 3.5]{712}). However, the extension of our arguments to (the shape dynamical version of) full general relativity and quantum mechanics is still a work in progress.
For current purposes, complexity is essentially the amount of clustering of a system, with a cluster being a set of particles that stay close relative to the extension of the total system. Next, we demand that the complexity function grow when (i) the number of clusters do, and (ii) the clusters become ever more pronounced, namely when the ratio between the extension of the clusters to the total extension of the system grow.

Given these premises, perhaps the simplest measure of complexity for the $N$-body system is 
\begin{equation}
\mathsf{Com}(q)=-\frac{1}{m_{\mathrm{tot}}^{5/2}}\sqrt{I_{\mathrm{cm}}}\,V_N=\frac{\ell_{\mathrm{rms}}}{\ell_{\mathrm{mhl}}}\,,
    \label{complexity}
\end{equation}

where $I_{\mathrm{cm}}$ is the centre-of-mass moment of inertia, $V_N$ is Newton's potential and $\ell_{\mathrm{rms}}$ and $\ell_{\mathrm{mhl}}$ account for the greatest and least inter-particle separations, respectively. Thus, their ratio, \eqref{complexity}, measures the extent to which particles are clustered. 
%It is simple to realize that the opposite of \eqref{complexity} is just the shape space version of $V_N$. Hence, apart from the mass factor, $-C(q)$ is usually called \emph{shape potential}.\footnote{To be precise, all PSD models have an associated shape potential, which is related to the measure of complexity defined in each particular context.}

%Now, for $E_{\mathrm{cm}}\ge 0$, $I_{\mathrm{cm}}$ is concave upwards as a function of Newtonian time, and its time derivative (essentially, the so-called \emph{dilatational momentum}) is monotonic\footnote{The dilatational momentum is defined as $D\equiv\sum _a^N \mathbf{r}_a^{\mathrm{cm}}\,\cdot \mathbf{p}^a_{\mathrm{cm}}$ and plays a major role in the $N$-body system, as the Janus Point corresponds to $D=0$. Also, given its monotonicity, the ratio $D/D_0$, with $D_0$ some arbitrary choice, has arguably been used as a physical time variable (\citealp{529,706}).}, which further implies that $I_{\mathrm{cm}}$ is U-shaped, with a unique minimum that divides all solutions in half . This is the so-called \emph{Janus Point}:\footnote{This designation was first put forward in \cite{529}.} 
The important thing about the complexity function \eqref{complexity} is that it has a minimum (dubbed \emph{Janus Point})
%near this point 
and grows in either direction away from it (\citealp{529,706}).
%This dynamically defines a direction of increasing complexity, measured by the complexity function \eqref{complexity}, which, we recall, by construction, tends to grow secularly. 
It can be argued that this direction of increasing complexity be identified with the so-called \emph{arrow of time} for internal observers (ones within one of the two branches at either side of the Janus Point). Thus, we arrive at a description of the experienced arrow of time in terms of purely intrinsic properties of the unparametrized curve on shape space.

Not only is the experienced arrow of time, or duration, accounted for by means of the curve in shape space, so too is length. This is achieved by the so-called \emph{ephemeris equations}.\footnote{Ephemeris equations are heavily model-dependent. Explicit expressions exist for the $N$-body system (\citealp{729}) and Bianchi IX cosmological model (\citealp{730}).} For the purposes of this discussion, we need only point out the general expression:

\begin{equation}
    F_i=G_i(q^a, \alpha _I^a),
    \label{ephemeris}
\end{equation}
where $F_i$ stands for a function of either length or duration ($i=1,2$) and $G_i$ is a function that solely depends on intrinsic properties of the curve, $\{q^a, \alpha _I^a\}$. Hence, the PSD framework promises to deliver a general formal mechanism for the appearance of ``everyday'' spatial and temporal structures from the fundamentally unparametrized dynamics of the the theory. 

Finally, let us consider how PSD recovers the standard dynamics for subsystems of the universe. In the classical $N$-body case, the dynamics features generic solutions which break up the original system into subsystems, consisting of individual particles and clusters, that become increasingly isolated in the asymptotic regime \citep{721}. Such almost isolated subsystems will develop approximately conserved charges, namely the energy $E$, linear momentum $\bf P$ and angular momentum $\bf J$. Within the dynamically formed subsystems, there are pairs of particles that may function as physical rods and clocks. These are referred to as \emph{Kepler pairs} because their asymptotic dynamics tends to elliptical Keplerian motion.

In conclusion, PSD represents a natural evolution of Barbour and Bertotti's original ideas, which provides a robust formal framework for a full ``Machianization'' of physics. Of course, much work still has to be done but, as the previous discussion shows, the technical foundations of the theory are already laid down.

\section{A Fascinating Metaphysical Diversity}\label{sec:3}
In the previous section, we have seen the fundamental theoretical elements that enter the dynamical description of a system according to PSD. Simply speaking, what is strictly needed to start constructing this dynamical description amounts to two elements: The shape space of the system and the dynamical law(s) \eqref{curve0}. The former represents the topologically structured collection of all constructible intrinsic configurations $q^a$ of a given system, while the latter generates the dynamical curve of the system, which is nothing but an ordered series of relational configurations.

For a scientific realist, the above description represents only half of the story. What is missing at this point is the characterization of a metaphysical link between the formalism of PSD and the physical world. Otherwise said, if we believe that PSD tells us something about the nature of the world, we should point out what it is that the formalism refers to. In order to address this point, two distinct but deeply inter-related questions should be asked: \emph{What} there is in the world, and \emph{how} the world is according to PSD. The first question concerns the \emph{ontology} of the theory, while the second concerns the best \emph{metaphysics} for the physical world compatible with the framework. We take this distinction as a mere working hypothesis, glossing over the actual relation between ontology and metaphysics---we will just assume that metaphysics does not (entirely) depend on ontology. This is, in fact, a controversial matter, largely orthogonal to our discussion (see, e.g., \citealp{728}, who instead defends the conceptual priority of ontology over metaphysics).

The question regarding the ontology of PSD is rather complex and tricky. Indeed, the ontological characterization of PSD depends on the actual implementation of the quotienting out procedure presented in the previous section and, hence, on the features of the starting theory subjected to such a procedure. For example, in the case of an $N$-particle Newtonian model, the quotienting out procedure ``transforms'' an ontology of material particles inhabiting a Newtonian background space into one of a ``web'' of particles related by Euclidean spatial relations (i.e., a Newtonian shape). The same applies to the general relativistic case, where an ontology of conformal $3$-geometries is arrived at once the quotienting out procedure is performed on the space of Riemannian $3$-geometries (see \citealp{427} for a comparison of the ontologies of SD and general relativity). From this point of view, the quotienting out procedure can be seen as a \emph{reconceptualization} that renders a starting fundamental ontology fully Leibnizian/Machian.

The above discussion begs the question as to whether PSD's ontology is model-dependent, i.e., whether we should ask ontological questions only in the context of the particular model considered. A positive answer to this question would point at some sort of perspectival realist attitude, in the sense that it would imply that PSD cannot deliver a unique and objectively true ``fundamental inventory'' of the world: Each model has ``its own'' ontology, so to speak (see, e.g., \citealp{630}, for a characterization of perspectival realism). Such an attitude, however, may be seen as dramatically weakening one of the strongest motivations for pursuing the PSD program, i.e., to provide a genuinely unifying framework for physics---and not just an unrelated collection of ``reconceptualized'' physical models. 

%\footnote{This may seem reminiscent of Carnap's ``internal questions''. Taking this conceptual resemblance too seriously, however, would lead in a direction that is arguably opposite to that, which a genuine scientific realist would follow.}

In order to deny that the ontology of PSD is model-dependent, two strategies are available. The first is to take all the models at once as providing the ``ontological bedrock'' of the world. This choice does not seem ideal since it would inflate the fundamental ontology while simultaneously depriving it of explanatory power (e.g., no explanation of the fact that Newtonian gravitation is an approximation of general relativity). The second option is instead to argue that there is in fact a class of models that faithfully represent the actual world (in the sense of being fully empirically adequate), while the remaining models should be seen either as physical \emph{possibilia} or as providing ontologies that are derivative on (in the sense of supervening on, or even being reducible to) the fundamental one. Such a choice is ontologically more parsimonious, and has the advantage of explaining how different physical domains are related by way of approximations or limiting procedures. Of course, this option will become available only if, and when, a full characterization of PSD will be available, which will include classical, quantum, and quantum-gravitational physics. Moreover, a consistent characterization of the dependence relations between the ``fundamental'' and the ``derivative'' models should be provided.

It is now clear why answering the ``What is there?'' question is not straightforward in the context of PSD, given the work-in-progress status of the framework. Hence, we leave it at that for the time being. In the following, we will just take for granted that a shape is a configuration of \emph{something}, without inquiring further into what this something is. We will instead inquire into the type of metaphysics that is best suited to characterize a world made of shapes behaving according to \eqref{curve0}, i.e., we will consider some possible answers to the question ``How is the world, according to PSD?''

\subsection{Shape Space Realism}\label{sec:srealism}

Let's start from what is probably the most straightforward metaphysical reading of the theoretical framework of PSD, that is, \emph{shape space realism}. According to this reading, \eqref{curve0} is just an algorithm that singles out a curve in shape space when it is fed certain initial conditions $\{q_0^a,\alpha _{0I}^a\}$. However, there is nothing ontologically significant that privileges some initial conditions over others; ditto for the ensuing curves. In other words, all points (and sequences thereof) in shape space have to be taken ontologically on a par. Let's unpack this last statement. First, we have to spell out what a point in shape space represents, then we have to specify in what sense these ``things'' must be considered equal from an ontic perspective. 

The first task is quite easy to accomplish, recalling the quotienting out procedure discussed in the previous section. The outcome of such a procedure is to strip a universal spatial configuration of elements of reality---that is, a universal snapshot of material particles, or field magnitudes, or anything else the original theory is about---of their unobservable degrees of freedom that are usually associated with the symmetries of the system. This is the way the starting non-relational configuration space of the system is translated into the fully intrinsic configuration space that we call shape space. Hence, each point in shape space represents an ``instantaneous'' universal relational configuration of elements of reality. For example, starting from a Newtonian $N$-particle system, we get a space where each point represents a configuration of $N$ particles related through the Euclidean spatial relations that survive the quotient by the similarity group. In this case, shape space is the structured set of all the possible shapes that can be constructed with these ingredients. This idea can be easily generalized, e.g., to the general relativistic case. In this case, shape space is the collection of all Riemannian $3$-geometries invariant under spatial diffeomorphisms and conformal transformations.\footnote{In principle, the same reasoning can be applied, \emph{mutatis mutandis}, to quantum theories. Recall from the discussion of tenet \ref{MachP3}, that the quotienting out procedure works whenever the starting theory exhibits a symmetry group $\mathcal{G}$ whose elements represent redundant degrees of freedom. There is no reason to doubt that this works in the quantum domain as well. For example, in the case of the de Broglie-Bohm theory, we expect the corresponding PSD models to be very similar to the Newtonian ones given that both theories are about material particles inhabiting a background Newtonian space (the quantum motion on shape space will substantially differ from the classical one, obviously). The fate of the wave function in PSD is an entirely different issue (see section \ref{sec:4}).}

Now that we have clarified what shape space is ``made of,'' we should take up the second task, i.e., specifying the sense in which all shapes featuring in this space are ontologically on a par. One may be tempted to say that shape space is the collection of all (physically) \emph{possible} shapes (according to the relational ingredients inherited from the quotienting out procedure). This modal characterization,\footnote{To simplify the discussion, we are subsuming modal talk under ``possible worlds'' parlance, being aware that the notion of ``possibility'' and that of ``possible world'' may not coincide, depending on the particular metaphysics of modality adopted (see, e.g., \citealp[][pp. 230-232]{284}, on this score).} however, would introduce a possible source of ontological asymmetry in that it may be argued that a subset of these shapes must represent the \emph{actual} world, that is, the one curve in shape space which is generated when the initial conditions obtaining at our world are fed into \eqref{curve0}. The way out of this challenge is to claim that in shape space there is no ``actual'' configuration as opposed to a ``possible'' one. The dynamics of the theory makes it the case that we get to experience some configurations rather than others, but this has nothing to do with the reality of each of them. In other words, all shapes are actual. In this way, shape space becomes the maximal collection of all the universal relational configurations \emph{simpliciter}. Otherwise said, for shape space realists, shape space \emph{is} the actual world.\footnote{Note that this kind of proposal goes back to the early stages of the Machian program, even before SD was born. Indeed, \cite{136,10} proposed to call the substantival relational configuration space ``Platonia,'' and the configurations with higher probability to be experienced ``time capsules.''} This, of course, does not mean that shape space is some sort of ``necessary being.'' Indeed, a shape space realist does not deny that the structure of shape space might have been different, which means that such a different shape space represents a possible world.

We have finally reached the metaphysical picture of the physical world according to shape space realism: All the universal ``snapshots'' simply exist all at once in a radically timeless and changeless sense. The dynamics encoded in \eqref{curve0}, and the ensuing constructions \eqref{complexity} and \eqref{ephemeris}, serve the only purpose of explaining the illusion of there being ``time'' and ``change'' in terms of the formation of stable records---including mental records---in an unparametrized sequence of configurations generated by certain (arbitrary) initial conditions. In this way, shape space realism is able to provide a straightforward metaphysical link between the theory's formalism and this timeless and changeless physical world. Simply speaking, all the mathematical structures entering PSD dynamics directly refer to geometric features of shape space \emph{qua} physical space. In this sense, shape space realism can be seen as a close cousin of the configuration space realism proposed in the context of quantum mechanics (see, e.g., \citealp{688}).

As straightforward as it may be, shape space realism as a metaphysical stance is far from being a naive metaphysical reading of the PSD formalism. In fact, it exhibits some peculiar features that are not usually ascribed to relationalism. First of all, it is at the same time a radically anti-realist (even idealist) and a radically realist position. It obviously denies the reality of much of what we usually call ``manifest image'' of the world, such as the passing of time, change---including motion in ordinary space---, but also the notion of identity of an object---which is usually intended as identity \emph{over time}. At the same time, however, it affirms the existence of each and every relational configuration compatible with the quotienting out procedure. The way it supports this ``ontological democracy,'' so to speak, is by claiming that these configurations are parts of a structured substance. Such a substance bears topological as well as metrical properties and, hence, it is some sort of ``space'' akin to the $3$-dimensional physical space we are acquainted with---even though it is much more structured than a simple Euclidean space.

This is possibly the most surprising feature of this stance. Although the whole starting point is the usual relationalist one, the end result is a position that is undoubtedly realist with respect to \emph{some sort} of space. One may retort that such a space is not a collection of ``points'' in the usual geometric sense but, rather, it is ``made of'' relational configurations in the plain metaphysical sense of ``being composed by.''\footnote{Another option may be to argue the other way around, i.e., things being shapes in virtue of them being points of shape space. Such an option, if defensible, would constitute a supersubstantivalist take on relational physics (see \citealp{718}, for a recent review of supersubstantivalism). We will set aside this possibility since, ironically, there is not enough space to discuss it here.} However, this would be a partial truth. In fact, composing a space out of configurations requires not only that said configurations ``be there,'' but also that they be ordered in a precise topological sense. The shape space realist cannot explain such a structure using facts inhering into the configurations themselves. The shapes are topologically ordered as a primitive fact of the matter, and this primitive topological structure represents the nature of shape space itself.

An immediate objection comes to mind at this point. Isn't it the case that shape space displays the hallmark of an absolute structure? After all, it is an immutable structure onto which we, for some reason, happen to project the illusory image of an ever-changing $3$-dimensional world. This means that, as for the case of Newtonian absolute space, shape space is just there, indifferent to the physics encoded in \eqref{curve0}. It is easy to realize, however, that this analogy cuts no ice. While in fact Newtonian space is an arena where material facts happen, in shape space nothing really happens: This space is some sort of ``block universe'' where all physically allowed universal snapshots are given all at once, timelessly. So, even if it is true that no physical happening can influence shape space, this is just because in the shape space realist picture there is no physical happening at all.

The above discussion prompts a second and more compelling objection. The starting point for this objection is that, usually, the point of contact between theoretical predictions and empirical observations can be traced back to material objects inhabiting physical $3$-space at a certain time. This is, after all, the kind of ``stuff'' labs are made of. Hence, in order to be empirically testable, a theory should provide a coherent story that links some theoretical terms to observable objects localized in ordinary space and time. The objection can hence be articulated as follows: How does shape space realism account for empirical observations, given that it regards ordinary space and time as some sort of illusion? This is, perhaps not surprisingly, a problem analogous to that encountered by configuration space realists (see, e.g., \citealp{585}, chapter 4, for an illuminating discussion of the issue at stake). This is not to say that PSD \emph{qua} theoretical framework does not have the resources to provide a technical account of such a story---recall again the mechanism behind equations \eqref{complexity} and \eqref{ephemeris}. However, adherence to shape space realism strips these technical reconstructions of ordinary space and time of metaphysical significance, raising the question of how we get to perceive these illusory constructs in the first place. However, at the present stage, this remains an open problem for those willing to defend shape space realism.

\subsection{Shape Monism}

One may object that shape space realism is not, in fact, a straightforward metaphysical reading of the PSD framework. This objection is based on the fact that the characteristic physical traits of PSD are not to be traced back to the quotienting out procedure that leads to shape space, but to the dynamical laws \eqref{curve0}. Hence, it makes more sense to be realist towards the curves generated by \eqref{curve0}, rather than to shape space as a whole. But what does this possibly mean? How should we conceive of a curve in shape space as (part of) the physical world?

The answer to this question involves a two-step argumentative strategy similar to that discussed at the beginning of subsection \ref{sec:srealism}. First, a characterization of a shape is called for. Here it is possible to repeat what was already argued for above, i.e., that shapes are ``instantaneous'' universal relational configurations of elements of reality represented by points in shape space. Curves in shape space, then, are a mathematical depiction of a dynamical sequence of shapes.

Secondly, and differently from the shape space realism case, it has to be argued that not all points (or curves) in shape space are to be taken ontologically on a par. Such a move implies that the physical world is a much more constrained structure than a vast block of kinematically allowed shapes. The world-building constraint is, in fact, \eqref{curve0}. The effect of this tenet is to deconstruct shape space into a myriad of worlds represented by the integral curves of \eqref{curve0}, each of which generated by a particular set of initial conditions. Among this huge set of possible worlds, the actual one is trivially represented by the curve generated by the initial conditions actually obtaining (whatever they are).

The worlds generated by \eqref{curve0}, as already said, are much smaller and more ordered than shape space as a whole. They are basically a linear sequence of universal configurations, so that now the ``snapshot'' metaphor makes more sense: A solution of \eqref{curve0} is analogous to a film strip showing the entire history of a universe all at once. Note that, also under this view, there is nothing inherently temporal in the fundamental picture of reality. Given that the curves generated by \eqref{curve0} are unparametrized, there is no fundamental sense in which the ordering is directed: There is no forward or backward direction in which the film stock unwinds. What is fundamental is the sequence of shapes given all at once in a timeless sense. From this point, the challenge is to come up with an explanation of how we get to observe a universe where (directed) time and change play an important role. We will defer this task to future work, being content to note that, given the resources mentioned in subsection \ref{psd} (i.e., complexity function, ephemeris equations, asymptotic Keplerian motions), it is much easier for the ``film stock'' view of relational dynamics to recover the manifest image of the world, rather than the shape space realist.

Of course, just saying that a world is a film stock whose frames are shapes is not enough to metaphysically characterize such a world. Indeed, even before starting to reflect on how to recover time from the film stock, a story is needed about the extent to which the film stock depicts a \emph{unique} evolving subject and it is not just a collection of pictures of different and unrelated subjects. This is a compelling problem. Each shape, in fact, is a universal relational configuration of ``stuff,'' but it is not possible to claim just that each shape represents \emph{the same} stuff at different times: The dynamics of PSD lacks the resources to establish trans-shape identity in this way---recall that PSD does away with best-matching, which was a way to establish a notion of equilocality in a relational setting. Without a solution to this problem, it is difficult to conceive of a curve in shape space as an organic whole representing the evolution of a single world (possible or actual). So, at this point, shape space realism seems to have the upper hand in that it works without needing any notion of trans-shape identity whatsoever.

But, if the problem with trans-shape identity comes from the difficulty to establish the identity of the elements of a starting configuration throughout the stages of dynamical evolution, a first step towards solving this conundrum may be to challenge the very assumption that the identity of a shape comes from the stuff (\emph{relata} as well as relations) it is made of. If facts about the identity of a shape \emph{as a whole} come before---in some appropriate sense of ``before''---facts about the identity of its parts, then it seems \emph{prima facie} simpler to argue that a curve generated by \eqref{curve0} depicts a unique relationally evolving subject.

The most straightforward way to implement this strategy is to claim that a shape is not \emph{made of} stuff. In other words, the relations and \emph{relata} do not compose a shape, in the sense that they are not metaphysically prior to the whole represented by a shape. The notion of metaphysical priority can be best understood in terms of grounding: For example, Socrates is metaphysically prior to the singleton $\{\text{Socrates}\}$, because the existence of $\{\text{Socrates}\}$ is grounded in the existence of Socrates. Otherwise said, since grounding can be intended as a partial ordering with respect to fundamentality, ``being ontologically prior to'' basically means ``being more fundamental than.'' Hence, the proposed way out is to claim that the shape as a whole is a fundamental object, with its proper parts---represented by relations and \emph{relata}---being metaphysically dependent upon, or posterior to, it. This is, in a nutshell, the main claim of \emph{shape monism}.\footnote{The general characterization of monism provided here heavily relies on \cite{538}. The reader is invited to check this source for an exhaustive discussion of this position.}

The main justification for shape monism is evident: The very notion of shape is that of an organic, integrated, universal whole that does not need anything external to itself in order to be characterized. This justification is supplemented with the subversion of the standard mereological tenet that the whole depends on its parts. This is an important point to be reiterated: Shape monism does not claim that a shape has no parts, but that the whole is prior to its parts. Hence, the talk of subsystems of the universe makes perfect sense also under shape monism. Subsystems---even the most simple ones---are perfectly real entities which, however, are not fundamental (i.e., they metaphysically depend on the whole).

At first sight, shape monism may look bizarre from a relationalist perspective. After all, this type of monism is far from asserting the fundamentality of spatial relations. To the contrary, these relations are parasitic upon the whole shape. But is this really so embarrassing for the relationalist? Maybe this initial discomfort results unwarranted upon closer inspection. First, shape monism salvages the core of the relationalist doctrine, i.e., the denial of space and time as ontologically independent substances. Indeed, a shape is not something that is placed in an external space; rather, it ``weaves up'' a suitable embedding space through the appropriate local constructions \eqref{ephemeris}. Second, it may be argued that shape monism perfectly captures the deep interdependence of all the elements of a shape, in the sense that each and every single element should be characterized relationally, not intrinsically. \cite{77} beautifully summarizes this point as follows:

\begin{quote}
[L]eibniz and Mach suggest that if we want to get a true idea
of what a point of space-time is like we should look outward at the universe, not inward into some supposed amorphous treacle called the space-time manifold. The complete notion of a point of space-time in fact consists of the appearance of the entire universe as seen from that point. Copernicus did not convince people that the earth was moving by getting them to examine the earth but rather the heavens. Similarly, the reality of different points of space-time rests ultimately on the existence of different (coherently related) viewpoints of the universe as a whole.\\
(\emph{ibid.}, p.265)
\end{quote}

In short, all problems between monism and relationalism disappear once we realize that the former is not a thesis about the metaphysical priority of relations but, rather, about the relatedness of all the parts of the cosmos. To sum up, shape monism is a viable option for a metaphysics of PSD.

This is, however, part of the story. The problem remains as to how to make sense of a curve as a succession of dynamical stages of a \emph{single} shape. The most simple solution is to point out that the identification of all shapes as different configurations of a unique ``cosmic object'' is just assumed as a primitive fact. Recall that the quotienting out procedure discussed in subsection \ref{Mach} was introduced to strip a single, well-defined global system (e.g., a Newtonian $N$-particle system) of unobservable degrees of freedom. So, although the procedure deletes any ``label'' that \emph{relata} and relations had (since a fully relational theory should include permutations as one of the symmetries quotiented away, see footnote \ref{perm}), still it does not impact on the primitive identity of the whole object being rendered fully relational.

The argumentative strategy sketched above, obviously, needs further refinement. First of all, claiming that, as a brute fact of the matter, there is a ``universal object'' whose dynamical evolution is captured by an integral curve of \eqref{curve0} looks more like a trivialization of the problem of trans-shape identity rather than a real solution. Secondly, it is doubtful whether it is a solution of the problem \emph{at all}. Shape monism just asserts the metaphysical priority of the whole over its parts, but it does not say much about how the identity of these parts should be carried over a dynamical curve. This renders it difficult to trace the evolution of a subsystem of the universe: Couldn't it be the case that different dynamical stages of a shape represent different ``carvings'' of this shape into smaller parts? If that was the case, then the identity of any subsystem would be destroyed in the transition from one dynamical stage to the other.  This is not by itself a fatal flaw of the position, but just an indication that still a lot of work lies ahead for the shape monist.

\subsection{Shapes as Ontic Structures}

Shape monism is not the only way to take equation \eqref{curve0} metaphysically seriously. An immediate option suggests itself once we recall that \eqref{curve0} is more of an equation \emph{schema} that encompasses the ``dynamical gist'' of relational physics according to PSD, rather than a full-fledged dynamical law. In other words, it is necessary to specify the system under consideration in order to fill in the details of \eqref{curve0}. Obviously, \eqref{curve0} is not a general dynamical scheme by accident. To the contrary, it represents a compact description of what is common to the relational dynamics of any possible physical system that can undergo the quotienting out procedure. But what can these ``common traits'' be? To answer this question, let's consider the general issue stemming from theory change in physics.

In a nutshell, this issue arises whenever we want to give a realist understanding of a physical theory, and amounts to pointing out that such a physical theory is subject to scientific revision and may in the future be dramatically altered or even discarded. How can we take such a theory as telling us something compelling about the physical world if it may soon be replaced by a more empirically accurate theory? For example, an early 19th Century physicist may have entertained the idea that heat was in fact a substance called \emph{caloric}, but a modern physicist regards such an ontic commitment as obsolete and inaccurate under the light of modern statistical mechanics.\footnote{Here we are simplifying the discussion a lot. The interested reader can start to dig deeper into this issue in the philosophy of science by taking a look at \cite{674}.} Among the possible responses to this challenge, one of the most famous---and interesting in the context of this chapter---was given by Henri Poincar\'e (\citealp{675}, p. 15). 

According to Poincar\'e, physical theories are unable to successfully grasp the fundamental ontology of the natural world---what he calls the ``real objects''---, because this aspect of reality is unknowable to us (``eternally hidden from us by nature''). This, however, does not mean that we should regard physical theories as recipes to get empirical predictions, in a somewhat instrumentalist fashion. There is, according to Poincar\'e, something real about the physical world that empirically adequate theories are successful in capturing. These real features of the world are nothing but the relations between physical objects. So, for example, what 19th Century physicists called ``caloric'' can be identified with what we today call ``heat'' insofar as both designations are intended to refer to some intimate interactive processes underlying material substances. The gist of Poincar\'e's response is the core tenet of modern \emph{structural realism} in philosophy of science.\footnote{This is not to say that Poincar\'e was the only ``responsible'' for the rise of scientific structuralism. See \cite{676}, for the first articulation of this stance in a modern context, and \cite{719}, for an introduction to the debate.} Hence, for structuralists, it is not anymore a problem if there is a change in the theoretical entities postulated by physical theories, since they maintain that it is the way such entities are related that is preserved through theory change. The empirical successes of different theories are then explained by the fact that they capture some structural aspects of the world.

Having the structuralist response in mind, it is now easy to answer our starting question. \eqref{curve0} is a compact way to convey the relational aspects of the dynamics that are common to all possible systems to which the quotienting out procedure can be applied. Whatever the fundamental ontology of the physical world might be (material particles, field magnitudes, strings, or something even more exotic), the key relational aspects of the dynamics are already captured by \eqref{curve0}. These aspects are exactly the features of reality that physical theories like Newtonian mechanics and general relativity get right. Then, the fact that, say, general relativity is more empirically accurate than Newtonian mechanics can be traced back to the actual implementation of \eqref{curve0} in the two cases---the general relativistic implementation of \eqref{curve0} being more accurate than the Newtonian counterpart.

This certainly represents a step forward in giving a metaphysical reading of \eqref{curve0}, but it is not sufficient to do the job. Indeed, structural realism as briefly characterized above looks more like an epistemic thesis than a metaphysical one, i.e., it is about what we can know rather than how the world is. This impression is reinforced by Poincar\'e's skeptical attitude towards the possibility of fully metaphysically informing our knowledge of the world. Luckily enough, structural realism can be turned into an ontic stance by simply arguing that the world is in fact \emph{made of} structures \citep[see, e.g.,][]{535}. This response dissolves Poincar\'e's worry in that it denies that the fundamental ontology of the world is constituted by individual objects whose identity and features are \emph{intrinsic} to them. Instead, what there is is a ``web'' of relations and \emph{relata}, with the identity and features of the latter being determined by their ``position'' in the structure. The ontic structural realist may go even further by arguing that the \emph{relata} are entirely ontologically parasitic on the relations: Fundamentally, there are only relations. Here we will blur the distinction between the moderate and the radical versions of ontic structural realism, since this difference does not hinge on the conclusions we are going to draw. Hence, from now on, we will remain agnostic towards the degree of ontic dependence that \emph{relata} bear to relations in a structure.

This ontic variety of structural realism seems to be tailor-made for PSD.\footnote{Note, however, that structuralism does not mitigate the issues with establishing a proper ontology for PSD discussed at the beginning of the section. While, in fact, it washes away the need to settle for fundamental \emph{objects}, the problem is shifted to that of finding the fundamental \emph{relations}.} It is in fact easy to argue that a shape is a structure in this specific sense. Hence, under an ontic structural realist reading, \eqref{curve0} generates an ordered (i.e., structured) sequence of spatial structures. The challenge for the structuralist, from this point on, is to provide a sound metaphysical picture (relying on the constructions \eqref{complexity} and \eqref{ephemeris}) of how space, time, and subsystems bearing their identity over time appear out of the ``structure of structures'' described by \eqref{curve0}. Let's try to sketch a possible strategy that the structuralist can follow to this extent. 

First of all, the structuralist may repeat the reasoning that led the monist to claim that a solution of \eqref{curve0} given certain initial conditions represents the entire ``history'' of a physically possible world (the actual world being described by the solution generated by the initial conditions being realized). From this point of view, the structuralist can exploit the monist's metaphor of a dynamical curve being a film stock that depicts the different stages of evolution of a cosmic structure. At this point, the challenge arises again, as to how to connect together the different ``frames'' of the film, i.e., the different shapes making the curve. Contrary to the monist, the structuralist cannot invoke some sort of primitive identity of the cosmos as a unique \emph{individual} object. In the present case, the metaphysical accent is put on structures, and it is evident from the formalism of PSD that there is a plurality of them---whereas the monist may argue that they are just different ways to ``ontologically carve'' the cosmos, which remains a singular entity. 

A simple way out of this issue is to bite the bullet and accept that the shapes making up a curve in shape space are distinct entities, i.e., there are no primitive facts about trans-shape identity of relations and \emph{relata} holding at a world. This obviously does not mean that these facts cannot be shown to \emph{supervene} on the succession of shapes laid down by \eqref{curve0}. PSD has in fact the formal resources to back up this supervenience picture. For example, having in mind that \eqref{curve0} orders configurations in terms of their similarity, and considering the construction underlying \eqref{complexity} and \eqref{ephemeris}, it is possible to argue---perhaps in a Humean fashion, as done for example by \cite{422} in the case of Newtonian dynamics---that there is a privileged embedding of the dynamical succession of shapes into a $4$-dimensional Riemannian geometry. The different configurations would then become different spacelike slices of this embedding ``spacetime.'' From this point on, the usual picture of individual objects stretching over spacetime in continuous trajectories could be recovered, with the identity over time of such objects being provided exactly by the fact that they are temporal stages of these spatiotemporal trajectories. Note that nothing in this construction would be assumed \emph{a priori} or, worse, would enter the fundamental structural ontology. Indeed, this construction would be a mere descriptive tool to cast the totally relational dynamics encoded in \eqref{curve0} in more familiar terms. 

The strategy sketched above seems to be a viable solution to the issue of finding a trans-shape identity criterion, as well as finding a story for the emergence of everyday space and time from PSD's relational picture. As a matter of fact, the above-mentioned construction has already been proposed in the Newtonian as well as in the general relativistic cases (see \citealp{469,470}, as well as \citealp{684}, section 4), although a concrete implementation in the PSD case is still an open question. 

\section{Conclusion}\label{sec:4}
The previous sections highlight the tremendous progress that relationalism has undergone in the last few decades. It all started from some bold conjectures on how to technically implement Leibniz and Mach's ideas, and ended up in an equally bold program that seeks to subsume physics in its entirety under the relational flag. However, as the discussion in section \ref{sec:2} makes it clear, this program has still a long way to go. One of the most pressing issues, from this point of view, is finding a concrete implementation of quantum physics, including quantum gravity.\footnote{Rovelli's covariant loop quantum gravity \citep{727} is an already well-established quantum gravity program that is based on relational considerations. However, Rovelli's relationalism is markedly non-Machian in that it denies any ``global'' structure to the world. For Rovelli, the world is a constellation of unrelated events consisting of the local interactions between physical systems.} This task is formidable, but it is nonetheless within the PSD's program abilities. 

Given that PSD's main tenet is to describe dynamics solely in terms of the intrinsic properties of the unparametrized curve in shape space, it should be expected that the prominent feature of quantum dynamics, i.e., the linearly-evolving wave function, should be reduced to such intrinsic properties as well. Just to give a glimpse of the current line of research in this direction, we just point out that an effective reduction of the wave function to the geometric properties of the curve potentially requires that the family of \emph{all} dynamical curves, for all possible initial conditions, be considered \emph{at once}, unlike the single curve of classical dynamics. This is analogous to the way the path integral formulation of quantum mechanics generalizes the action principle of classical mechanics, i.e., by replacing the unique classical trajectory for a system with a multiplicity of quantum-mechanically possible trajectories. This ``path multiplicity'' clearly cannot be readily made reconcilable with the deterministic nature of the Mach-Poincar\'e principle. One may envisage, at this point, the rise of some relational counterparts of the interpretations of quantum mechanics, for example with respect to the ontic status of these trajectories (all existing in parallel, or just one of them being ``actualized''). And, perhaps, some benefits for the standard debate in quantum foundations could come from moving to a fully relational arena.

In conclusion, the road to quantum relational physics is full of hurdles, but it nonetheless points to an exciting and promising direction.

\section*{Acknowledgements}\pdfbookmark[1]{Acknowledgements}{acknowledgements}
We are very grateful to an anonymous referee and the guest editor Andrea Oldofredi for their enlightening comments on a previous version of this chapter. P.N. and A.V. acknowledge financial support from the Polish National Science Centre, grant nr. 2019/33/B/HS1/01772.

\pdfbookmark[1]{References}{references}
\bibliography{biblio}

\end{document}